\documentclass[epj,referee]{svjour}
\usepackage{amsfonts}
\usepackage{amssymb}
\usepackage{amsmath}
\usepackage{epsfig}
\usepackage{hyperref}
\begin{document}
\title{From crystal to amorphous: a novel route to unjamming in soft disk packings}
\author{Fabricio Q. Potiguar}
\institute{Departamento de F\'\i sica, ICEN, Av. Augusto Correa, 1, Guam\'a, 66075-110, Bel\'em, Par\'a, Brazil\\ \email{fqpotiguar@ufpa.br}}
\abstract{
Numerical studies on the unjamming packing fraction of bi- and 
polydisperse disk packings, which are generated through compression of a 
monodisperse crystal, are presented. In bidisperse systems, a fraction 
$f_+=0.400$ up to $0.800$ of the total number of particles have their radii 
increased by $\Delta R$, while the rest has their radii decreased by the same 
amount. Polydisperse 
packings are prepared by changing all particle radii according to a uniform 
distribution in the range $[-\Delta R,\Delta R]$. The results indicate that 
the critical 
packing fraction is never larger than the value for the initial monodisperse 
crystal, $\phi_0=\pi/\sqrt{12}$, and that the 
lowest value achieved is approximately the one for random close packing. These 
results are seen as a consequence 
of the interplay between the increase in small-small particle contacts and the 
local crystalline order provided by the large-large particle contacts.
}

\maketitle

\section{Introduction}
\label{intro}
The jammed state of condensed matter is characterized by the 
sudden arrest of a system's internal dynamics. Macroscopically, 
the jammed system develops an yield stress and behaves, essentially, like a 
solid. A pile of sand under the action of gravity, clogged flow of powders 
through a pipe or coagulated colloidal microstructures with high elastic moduli 
are a few examples with practical applications that exhibit jamming. On the 
theoretical side, random packings of hard and soft elements (spheres, disks, 
etc.) display similar behavior to those observed in real systems and are often 
used as prototype systems for the study of jamming. 
It has been extensively studied recently both by simulations 
\cite{Mak00,Her02,Her03,Zha05,Her05,Silbert05-01,Silbert06,Cia09-01,Cha09,Xu09} 
and experiments \cite{Che09-01,Che09-02}. 
As expected for strongly interacting systems, there are several theoretical 
proposals \cite{Fierro05,Henkes05,Hentschel07,Song08} to describe this state, 
most of them relying on approximate mean-field arguments. 

There are some facts that seem well settled today about jamming. 
First, a quench is an essential ingredient for a system to reach the jammed 
state. It prevents any crystallization that may occur during a slow 
rearrangement of particle positions during equilibration. Second, the critical 
jamming density is affected by 
particle size ratio \cite{Xu09}, shape \cite{Schreck10} and the preparation 
protocol, but its critical properties are the same (recently, Chaudhuri 
{\em et al}. \cite{Cha09} showed that this critical point is not unique, even 
for large 
systems). Third, the jamming point in monodisperse and bidisperse packings is 
manifested in structural properties as the $\delta$-function behavior of the 
first peak of the radial distribution 
function and the split of its second neighbor peak \cite{Silbert06}.


Most studies about jamming at zero temperature focus only in 
one preparation protocol (random packings of, possibly, 
overlapping particles, quenched to the nearest minimum energy state). 
One may ask, then, is it possible to produce a jammed state from a 
completely ordered system (crystal)? Is the quench alone enough to take the 
packing out of its global minimum of energy and trap it in a jammed state? In 
particular, if the jammed system prepared in this way will it be more dense 
than the initial crystal? All these question, and others, will be addressed in 
this paper. Jammed states will be produced by the quenching, and further 
decompression, of crystalline disk packings.


This initial condition, at first sight, is not suitable to produce a jammed 
state, since disordered packings are commonly associated with jamming. However, 
it will be shown that the structural features of jamming are present in such 
systems, hence regarding this preparation protocol as a valid one to produce 
jammed packings. It will be shown that this initial condition opens the 
possibility to reach jammed states in a distinct region of the 
packing phase diagram, which are unaccessible from ordinary, random initial 
packings algorithms. Along with the studies of bidisperse packings, the 
decompression of polydisperse packings is explored. 

Sect. \ref{sim} is where simulation details are provided, sect. \ref{results} 
holds all results and sect. \ref{summary} is reserved for conclusion.

\section{\label{sim}Simulation methods}
The particles are soft, elastic disks, which interact through linear springs. 
The compression potential energy between disks $i$ and $j$ given by:
\begin{equation}
\label{ener-law}
U_{ij}=\frac{1}{2}\kappa(R_i+R_j-r_{ij})^2\Theta(R_i+R_j-r_{ij}),
\end{equation}
where, $\kappa$ is the elastic constant (taken $\kappa=1$ and equal for all 
contacts), $R_i$ is the $i$-th disk radius, $r_{ij}$ is the distance between 
the disks, and $\Theta(x)$ is the step function. 

Initially, the packing consists in $N$ disks, of radius $R_0$ (taken as the 
unit length), arranged in a triangular lattice. The contact energy, eq. 
(\ref{ener-law}), is given in units of $\kappa R_0^2$. The system's periodic 
boundary lengths are given by $L_X=2R_0N_X$ and $L_Y=\sqrt{3}R_0N_Y$, where 
$N_X\times N_Y=N$, 
is the total number of disks. The values $N_X=N_Y=50$ are chosen, which gives 
$N=2500$, throughout the experiments. The results presented here are 
essentially the same for systems with $N_X=N_Y=10$ and $N=100$. This 
choice of boundary lengths perfectly accommodates a 
triangular lattice of equal disks, which implies that the initial 
packing fraction has the largest value for a two dimensional monodisperse 
system (the Kepler conjecture) \cite{Hal05}:
\begin{equation}
\label{pack-tri}
\phi_0=\frac{N\pi R_0^2}{L_XL_Y}=\frac{\pi}{\sqrt{12}}\approx0.907.
\end{equation}

\begin{figure}[h]
\rotatebox{0}{\epsfig{file=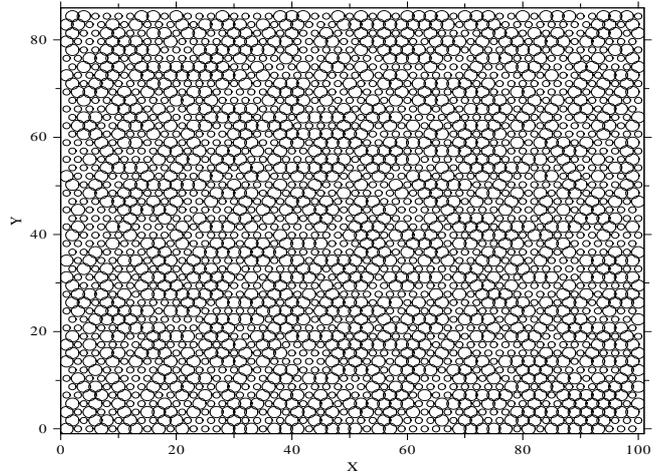,width=10.0cm,height=8.5cm}}
\caption{An illustration of a quenched packing ($f_+=0.500$ and 
$\Delta R/R_0=0.300$).
\label{quench}}
\end{figure}

The quench is performed by changing the disks radii by a suitable 
amount, the dispersity degree $\Delta R$. Such change is instantaneous, 
in order to trap the system in a jammed state (quench). For bidisperse 
packings, a 
number $N_+=f_+N$ of the disks (randomly chosen) have their radii increased by 
$\Delta R$ while the rest of the particles, $N_-=(1-f_+)N$, have their 
radii decreased by $\Delta R$ (similar to what was used in \cite{Boc92}). The 
number fractions chosen are $f_+=0.400$, $0.500$, $0.600$, $0.700$ and $0.800$. 
For polydisperse packings, the radii are changed by an amount uniformly 
distributed between $\left[-\Delta R, \Delta R\right]$. Fig. \ref{quench} 
shows a quenched configuration.
\vspace{2em}

These 
changes introduce a compression potential energy, since there will always be 
some overlap between nearest neighbor grown disks. 
Given the absence of energy dissipation, the minimization is performed by the 
conjugate gradient method \cite{nr}. A dissipative packing cannot be studied 
with this numerical method. A molecular dynamics (MD) approach is more 
suitable, and 
certainly will yield very distinct results (for a study on the phase diagram 
of dissipative packings see \cite{Cia09-02}). Finally, random packings can also 
be prepared with a MD approach by swelling void particles (void expansion 
method \cite{Schenker09}) that increase the volume fraction and takes the 
system through the jammed point.

It should noticed that a jammed state is not reached, with this protocol, at 
lower number fraction, for instance $f_+\leq0.300$.
Given the small probability to form large-large contacts at the beginning, 
the available space for rearrangements is larger than the one needed for 
complete relaxation. This leads to a {\em melting} of the packing, 
instead of jamming (the highlight emphasizes that the melting picture should 
be supported by additional simulations). 
Therefore, such low number fraction packings do not behave as their high $f_+$ 
counterparts. This will happen only if the mean radius of such packings is 
larger than $R_0$. 


Since the goal is to find the maximum packing fraction with a 
vanishing potential energy, an initial minimization is performed right after 
the quench. The energy 
minimum is achieved when the difference between the current and the last 
energy values is no more than $10^{-10}$.
After reaching the nearest energy minimum, particles are slowly decompressed, 
{\em i.e.}, particle's radii are decreased by a small constant amount $\gamma$ 
at each cycle, which provides a slightly larger space for them to relax 
(expansion step). After each expansion step, an energy minimization is 
performed in order to take the system closer to the zero potential energy 
state. The decompression is finished when the total potential energy is less 
than a predefined value, $\epsilon$. Hence, at the of the protocol for a 
suitable $\epsilon$ value, the system should be very close to the jamming point.
 In sect. \ref{res_params}, the influence of both parameters on the results 
will be shown.



The average packing fraction $\left<\phi\right>$, from now on referred to as 
the critical packing fraction (CPF), is measured at the end of the 
decompression as a function of $\Delta R/R_0$. The averages are over 
realizations (typically $20$ for each case) and the error bars are calculated 
as $\sqrt{\left<(\phi-\left<\phi\right>)^2\right>}$ in each case. 
Also, the packing structure and order are 
studied through the calculation of the Radial Distribution Function (RDF), 
$g(r)$, and the orientational order parameter: 
\begin{equation}
\label{psi}
\Psi_j=\frac{1}{z_j}\sum\limits_{k=1}^{z_j}e^{i6\theta_{jk}},
\end{equation}
where the sum runs over the $z_j$ nearest neighbors of disk $j$ and 
$\theta_{jk}$ is the angle between the line joining the $j$-th and $k$-th 
disks centers and the $x$ axis \cite{Sta97}. The absolute value of $\Psi_j$ is 
measured at the end of the full minimization, and is presented as an average 
over particles and runs. Its value is unity for a perfect 
triangular array of particle, while $\left|\Psi_j\right|<1$ for disordered 
packings. Two disks are considered first neighbors if they overlap.

It should be noticed that, in most studies of bidisperse packings 
\cite{Her02,Her03,Her05,Cha09,Xu09}, particle size differences are given in 
terms of the size ratio, $\sigma$, instead of the size difference, $\Delta R$. 
Both quantities are connected by:
\[
\frac{\Delta R}{R_0}=\frac{\sigma-1}{\sigma+1}.
\]



\section{\label{results}Results and discussion}
This sect. holds all numerical results. First, some results for the CPF as a 
function of the simulation parameters $\gamma$ and $\epsilon$ are shown.
Second, the full CPF results, along with the packing structure and order will 
follow, respectively.

\subsection{\label{res_params}Independence on the simulation parameters}
The simulation is controlled by two sensitive parameters, the decompression 
rate, $\gamma$, and the minimum compression energy, $\epsilon$. Figure 
\ref{rpf-params} holds the results for the CPF for three distinct parameter 
sets:

\vspace{2em}
\begin{figure}[h]
\rotatebox{0}{\epsfig{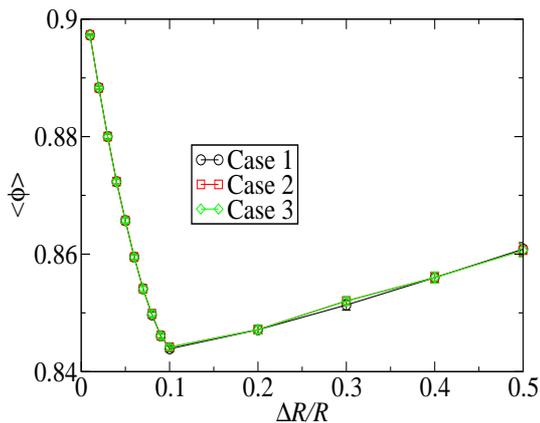}}
\caption{CPF for distinct values of the decompression parameter, $\gamma$, and 
the energy minimum, $\epsilon$.
\label{rpf-params}}
\end{figure}

The three cases presented, all for $f_+=0.500$, have the following set of 
parameter values: case 1, $\gamma=10^{-5}$ and $\epsilon=10^{-6}$; case 2, 
$\gamma=10^{-6}$ and $\epsilon=10^{-6}$; case 3, $\gamma=10^{-6}$ and 
$\epsilon=10^{-8}$. As seen in this figure, all results agree well within 
simulation error. Hence, all the following results will be given for case 1, 
unless noticed otherwise. This value of $\epsilon$ corresponds to 
a minimum energy per particle of the order $10^{-9}$. Similar results hold 
for the polydisperse packing (not shown).

\subsection{Jamming packing fraction}
\begin{figure}[h]
\rotatebox{0}{\epsfig{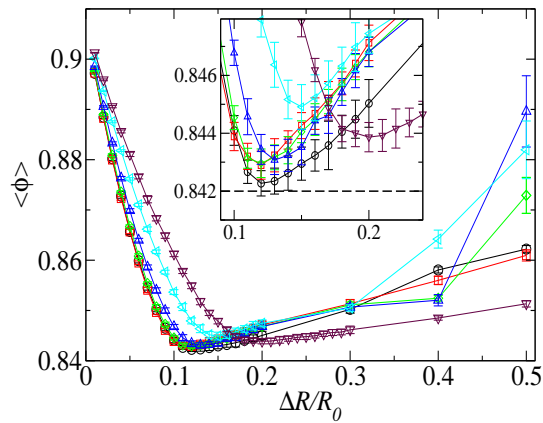}}
\caption{All results for the CPF, for distinct particle number fraction, $f_+$, 
and size dispersity. Symbols correspond to number fractions of $f_+=0.400$ 
(circles), $f_+=0.500$ (squares), $f_+=0.600$ (diamonds), $f_+=0.700$ 
(triangles), $f_+=0.800$ (left triangles) and polydisperse packing 
(inverted triangles). 
The inset is a zoom to the region close to the curves minima and the dashed 
line represents the CPF value for the RCP state \cite{Her02}, 
$\phi_{RCP}=0.842$.
\label{rpf-all}}
\end{figure}

In fig. \ref{rpf-all}, the results for the CPF for all 
number fraction, $f_+$, and size dispersity, $\Delta R/R_0$, values are shown.
\vspace{2em}

From this graph, one can see that the final packing fraction is 
never larger than the one for the triangular lattice, eq. (\ref{pack-tri}), 
and it goes through a minimum. From the inset of fig. \ref{rpf-all}, 
it can be seen that the minimum CPF changes with number fraction. Its lowest 
value is $0.843$ for $f_+=0.400$ at $\Delta R/R_0=0.120$, while the largest 
minimum is $0.845$ for $f_+=0.800$ at $\Delta R/R_0=0.150$. The CPF value for 
the random close packing (RCP) state is shown as a reference 
value, since it is the jamming packing fraction for the monodisperse packing 
\cite{Her02}. These results indicate that the RCP 
state corresponds to the lowest critical packing fraction achieved with this 
protocol, with packing properties distinct from the original study (in 
that case \cite{Her02}, $f_+=0.500$ and $\Delta R/R_0=1/6$). For the 
polydisperse packing, the minimum CPF is $0.844$ at $\Delta R/R_0=0.200$. 

Recent results on jamming of bidisperse sphere packings \cite{Xu09} and 
ellipsoid and dimer packings \cite{Schreck10}, that focused on the influence 
of the size ratio in the CPF, show that this quantity presents a maximum in 
the size ratio range $[1,\infty]$. These references also carried out their 
simulations with random initial packings.
(In \cite{Xu09}, the largest CPF occurs at $f_+=0.500$ and $\Delta R/R_0=1/3$). 
Here, all CPF values are also above the RCP one, but it goes 
through a minimum instead of a maximum. The reason behind this distinction seems
 to be mainly the initial packing. The jamming point in a random monodisperse 
packing may be increased with an appropriate size ratio (the system packs more 
efficiently). On the other hand, the jamming point for a regular packing can 
only be decreased from the value given in eq. (\ref{pack-tri}). This can be 
seen as a consequence of the fact that, as stated earlier, the triangular 
lattice is the most dense packing of equal disks \cite{Hal05}.
Therefore, one can conclude that performing a decompression simulation with 
a regular initial packing, one can reach a very dense jammed state, not 
accessible from a random initial packing. This is the main results of this 
paper and it is what is meant by a novel route to unjamming in the title. 

The small dispersity behavior of the CPF can be understood as follows. When 
the regular triangular array of disks is 
quenched, the total area occupied by the disks, initially given by 
$A_0=N\pi R_0^2$, is changed to: 
\[
A_{b0}=\pi\left[\sum\limits_{i=1}^{N_+}(R_0+\Delta R)^2+\sum\limits_{i=1}^{N_-}
(R_0-\Delta R)^2\right]-A_{ovlp}.
\]
where $A_{ovlp}$ represents the total overlapped area. Developing the squared 
terms and using the fact that $f_++f_-=1$, one reaches the following 
expression for the initial modified area:
\[
A_{b0}=N\pi R_0^2+2(2f_+-1)N\pi R_0\Delta R+N\pi\Delta R^2-A_{ovlp}.
\]
Since the experiment is carried through particle decompression at a constant 
rate, when 
the overlapped area vanishes, the total area occupied 
by the disks should be given by the exact same expression, but for a distinct 
mean radius $R$, instead of $R_0$:
\[
A_b=N\pi R^2+2(2f_+-1)N\pi R\Delta R+N\pi\Delta R^2.
\]
The value of this mean radius is $R=R_0-r$, with $r=\gamma n$, where $n$ is 
the number of decompression steps. Defining $x=\Delta R/R_0$ and $y=r/R_0$, 
and dividing both sides by $L_XL_Y$ one has:
\[
\frac{A_b}{L_XL_Y}=
\]
\[
\frac{N\pi}{L_XL_Y}\left[(R_0-\gamma n)^2+
2(2f_+-1)(R_0-\gamma n)\Delta R+\Delta R^2\right].
\]
The final average CPF, among different experiments with fixed 
boundary lengths can be obtained using the average $r$ value in 
$A_b$. Therefore, using eq. (\ref{pack-tri}), this relationship yields:
\[
\left<\phi_b\right>=
\]
\begin{equation}
\label{avg-rpf}
\phi_0\left[1+2(2f_+-1)x+x^2-2(2f_+-1)x\left<y\right>-2\left<y\right>+\left<y^2\right>\right].
\end{equation}

\begin{figure}[h]
\rotatebox{0}{\epsfig{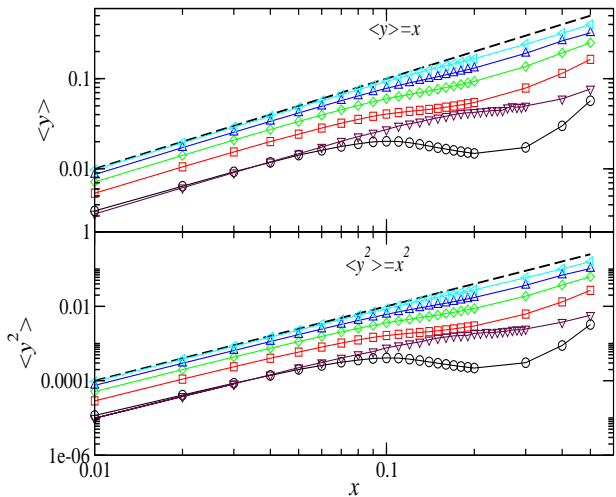}}
\caption{Upper panel: average number of decompression cycles. Symbols follows 
convention in fig. \ref{rpf-all}. Lower panel: average square number of 
decompression cycles.
\label{avg-rad}}
\end{figure}

A similar argument holds for a polydisperse packing. In this case, the area 
occupied by the disks after the initial compression is given by:
\[
A_{p0}=\pi\sum_{i=1}^{N}(R_0+\Delta R_i)^2-A_{ovlp},
\]
where $\Delta R_i$ is the change in the $i$-th disk radius. At the end of the 
decompression phase, the total disk area is:
\[
A_p=\pi\sum_{i=1}^{N}(R+\Delta R_i)^2,
\]
with $R$ given as above. Since the quantity $\Delta R_i$ is uniformly 
distributed in the range $[-\Delta R,\Delta R]$, one may use the moments of 
$\Delta R_i$,
\begin{eqnarray}
\label{mom-uni}
\frac{1}{N}\sum\limits_{i=1}^{N}\Delta R_i=0,\\
\label{mom-uni-2}
\frac{1}{N}\sum\limits_{i=1}^{N}\Delta R_i^2=\frac{\Delta R^2}{3},
\end{eqnarray}
to write
\[
A_p=N\pi\left[(R_0-\gamma n)^2+\frac{\Delta R^2}{3}\right].
\] 
Following the same steps as in the bidisperse case, the average CPF can 
be written as:
\begin{equation}
\label{avg-rpf-poly}
\left<\phi_p\right>=\phi_0\left[1-2\left<y\right>+\left<y^2\right>+
\frac{x^2}{3}\right].
\end{equation}
Hence, with the knowledge of the average number of decompression steps, one 
can match equations (\ref{avg-rpf}) and (\ref{avg-rpf-poly}) with the results 
given in fig. \ref{rpf-all}. The 
results for $\left<y\right>$ and $\left<y^2\right>$ are given in fig. 
\ref{avg-rad}.
\vspace{2em}


Consider the following argument to explain this result. 
If this decompression experiment was performed in the 
monodisperse case, where each disk has exactly $6$ contacts, simetrically 
placed around its center, no change in the structure will ever occur due to 
the decompression, and the packing fraction value in which the 
compression energy is zero would be given by (\ref{pack-tri}). Therefore, the 
disk radii should return to their original value in order to reach this packing 
fraction. This implies that $\left<y^\gamma\right>=x^\gamma$. Since at 
low dispersity this relationship is approximately realized, one can infer that 
at the jamming point, one has $\left<y\right>=ax^\alpha$ and 
$\left<y^2\right>=bx^\beta$. 
These four parameters $a,b,\alpha,\beta$ represent the effect of structure 
rearrangements in the quantity $r$. Table \ref{tab_1} holds their values as 
measured from power law fits to the curves in fig. \ref{avg-rad}. Since 
deviation from power law behavior do not occur at the same dispersity for all 
cases, the fits were performed up to $x=0.040$ for $f_+=0.400$, $0.050$ for 
$f_+=0.500$, $0.060$ for $f_+=0.600$ and $x=0.100$ for the other cases. The 
parameters approach their monodisperse values as one 
increases the number fraction. The polydisperse packing is an exception since 
it has no monodisperse limit. Also, one can see that, at all number fractions 
and the polydisperse case, the relationship $\beta=2\alpha$ holds. The data 
do not imply a simple relationship between the coefficients $a$ and $b$.\\

\begin{table}
\caption{Decompression step parameters, measured from the curves in fig. \ref{avg-rad}.}
\label{tab_1}
\begin{tabular}{|c|c|c|c|c|c|c|}
\hline
$f_+$ & $0.400$ & $0.500$ & $0.600$ & $0.700$ & $0.800$ & poly \\
\hline
$a$        & $0.214$    & $0.413$    & $0.588$    & $0.722$    & $0.895$ & $0.244$ \\
\hline
$b$        & $0.046$    & $0.171$    & $0.346$    & $0.520$    & $0.802$ & $0.061$ \\
\hline
$\alpha$   & $0.898$    & $0.941$    & $0.956$    & $0.956$    & $0.979$ & $0.941$ \\
\hline
$\beta$    & $1.80$     & $1.88$     & $1.91$     & $1.91$     & $1.96$ & $1.89$ \\
\hline
\end{tabular}

\end{table}

\begin{figure}[h]
\rotatebox{0}{\epsfig{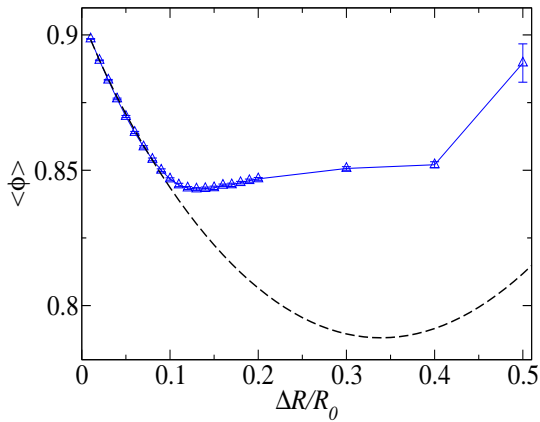}}
\caption{Eq. (\ref{avg-rpf-2}) plotted with the parameters correspondent to 
the $f_+=0.700$ case.
\label{fp70-fit}}
\end{figure}

By using this form for the average number of decompression steps, the average 
CPF, eq. (\ref{avg-rpf}), can be cast in the following form:
\begin{equation}
\label{avg-rpf-2}
\frac{\left<\phi\right>_b}{\phi_0}=1+2(2f_+-1)x+x^2-2(2f_+-1)ax^{\alpha+1}-2ax^\alpha+bx^{2\alpha}.
\end{equation}
This curve for the parameters $a$, $b$, 
$\alpha$ and $\beta$ given for the $f_+=0.700$ case is shown as a dashed line 
in fig. \ref{fp70-fit}. One can see that, at low dispersity, this eq. agrees 
well with the results. 
\vspace{2em}

For the polydisperse packing, the average CPF can be written as:
\begin{equation}
\label{avg-rpf-poly-2}
\frac{\left<\phi\right>_p}{\phi_0}=1-2ax^\alpha+bx^{2\alpha}+\frac{1}{3}x^2.
\end{equation}

The comparison between figs. \ref{rpf-all} and \ref{avg-rad} shows that the 
deviation from the power law behavior of $\left<y\right>$ coincides with the 
region where the CPF reaches its minimum value.
Since a complete understanding 
of this fact would be given by a more detailed (and complicated) theoretical 
approach, such as the one proposed in \cite{Song08}, only the packing 
structure is probed here.
The reason behind this choice is that the structure relaxation during 
decompression is the main cause for the results seen in figs. \ref{rpf-all} and 
\ref{avg-rad}.

\subsection{Packing structure}
\begin{figure}[h]
\rotatebox{0}{\epsfig{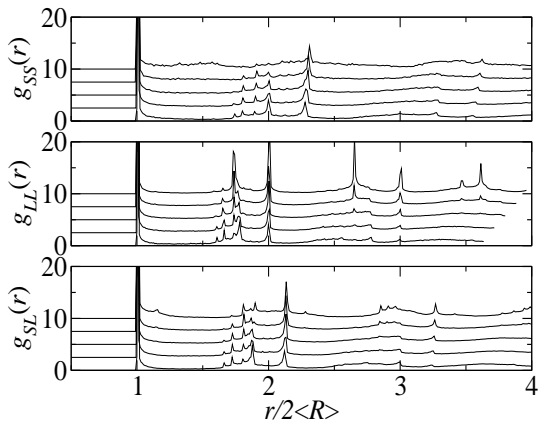}}
\caption{RDFs for a dispersity value of $0.120$. Each line, shifted for 
clarity, corresponds to a distinct number fraction (from top to bottom): 
$f_+=0.800$, $0.700$, $0.600$, $0.500$, $0.400$. 
\label{rdf-1}}
\end{figure}

\begin{figure}[h]
\rotatebox{0}{\epsfig{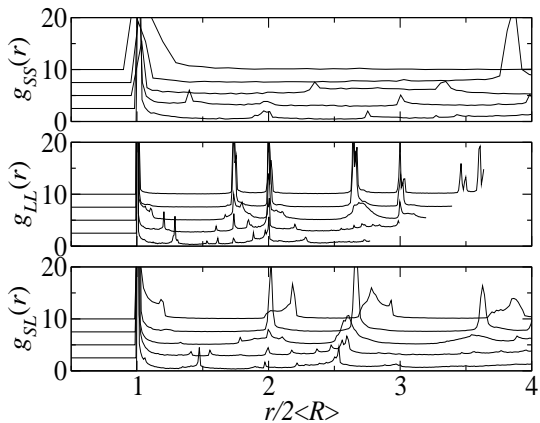}}
\caption{RDFs for a dispersity value of $0.500$. Curves are shown in accordance 
to the convention in fig. \ref{rdf-1}.
\label{rdf-2}}
\end{figure}

The RDF is measured at the end of the 
full minimization process. These measurements are performed regarding 
the type of particle contact, {\em i.e.}, the probabilities of 
small-small, $g_{SS}(r)$, large-large, $g_{LL}(r)$, and small-large, 
$g_{SL}(r)$, contacts are measured. In the triangular array of monodisperse 
disks, one expects $g(r)$ to have sharp peaks at $r=1$, $\sqrt{3}$ and $2$ 
particle diameters.
Figure \ref{rdf-1} shows all RDFs for 
$\Delta R/R_0=0.120$ and all number fractions. Figure 
\ref{rdf-2} has these same functions but for a dispersity value of $0.500$. In 
both cases, 
all interparticle distances are normalized by the corresponding final average 
particle diameter. For instance, for a small-small contact, the final small 
particle diameter is $2\left<R\right>=2(R_0-\Delta R-\left<r\right>)$, where 
$\left<r\right>$ is the average number of decompression steps, fig. 
\ref{avg-rad}. Hence, the distance range where these contact probabilities are 
measured appear distinct for each contact type and number fraction, especially 
for large-large contacts, which have a larger mean particle diameter.
\vspace{2em}

In all curves in fig. \ref{rdf-1}, the jamming 
structural signature, a delta-like first peak and a split second peak, are 
seen. Also, the second small-small contact peaks are not located precisely at 
$r=\sqrt{3}$ and $2$. Instead, they are shifted to the right, consistent with 
the results shown in \cite{Xu09}. A brief explanation of this fact is that 
these two peaks occur only when three particles form a triangular cluster. 
Hence, this type of cluster should be absent for small-small and small-large 
contacts. Moreover, small-large 
contacts have a second peak at intermediate positions between small-small and 
large-large contacts, as expected, since the mean diameter of such a contact 
is given by $2(R_0-\left<r\right>)$. 
\vspace{2em}

A marked feature of these graphs is that the $g_{SS}(r)$ and $g_{SL}(r)$ peaks 
between $r=1$ and $\sqrt{3}$ decrease with $f_+$, while those for $r=2$ become 
sharper. This indicates that, for larger $f_+$, small particle relax to 
positions farther away from each other, even though they start all at the same 
structure.

The $g_{LL}(r) $ peaks at $r=1$, $\sqrt{3}$ and $2$, increase and become 
sharper, while the intermediate ones between $r=1$ and 
$\sqrt{3}$ almost disappear at the largest number fraction. Since sharp peaks 
at $\sqrt{3}$ and $2$ are a signature of a triangular lattice structure, one 
may infer that such large particle rich packings relax to structures similar 
to a crystalline one. This fact also explains why the large-large peaks 
between $r=1$ and $\sqrt{3}$ become smaller. Such characteristic should 
be expected, since the initial packing is regular and only large-large particle 
contacts, at the outset, imply compression. Therefore, if a group of first 
neighbors become large particles, they will probably remain in this cluster 
up to the end of the process. 

\begin{figure}[h]
\rotatebox{0}{\epsfig{file=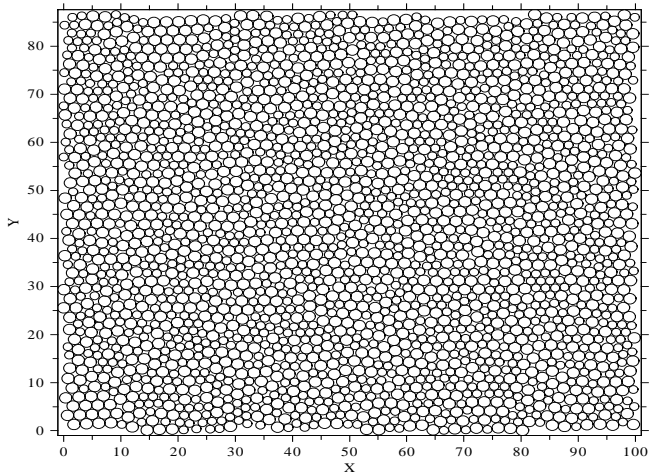,width=10.0cm,height=8.5cm}}
\caption{Packing configuration for $\Delta R/R_0=0.12$ and $f_+=0.50$. Particle 
periodic images are omitted.
\label{config_1}}
\end{figure}

An illustration of such a configuration is given in fig. \ref{config_1}. It 
shows a well mixed packing with an occasional pocket of large particle 
crystals (lower left corner). 
\vspace{2em}

\begin{figure}[h]
\rotatebox{0}{\epsfig{file=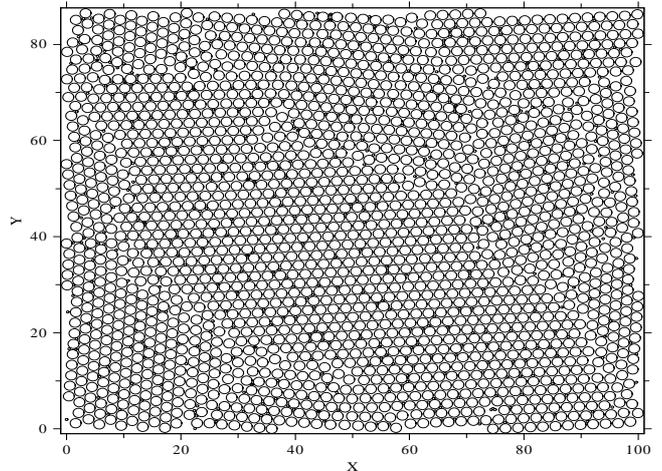,width=10.0cm,height=8.5cm}}
\caption{Packing configuration for $\Delta R/R_0=0.500$ and $f_+=0.800$. Small 
particles are enlarged for better visualization.
\label{config_2}}
\end{figure}

On the other hand, the pair correlation functions at large dispersity, fig. 
\ref{rdf-2}, shows markedly distinct features of the packing structure. 
First of all, small particle aggregates become 
progressively more distant, in small particle mean diameter units, for larger 
$f_+$. Second, $g_{LL}(r)$ at $f_+=0.400$ and $f_+=0.500$ shows 
several small peaks between those at $r=1$ and $2$, as in fig. 
\ref{rdf-1}, but the one at $r=\sqrt{3}$ is not easy to distinguish. Only at 
high number fractions this peak becomes clear. This means that large 
particles form, again, structures close to crystalline ones at high number 
fraction. Finally, one can see a clear change in $g_{SL}(r)$ 
from $f_+=0.700$ to $0.800$. The peaks for the lower number fraction are 
sharper than the corresponding ones at larger number fraction. This is 
probably due to the fact that small particles can be more easily accommodated in 
vacancies among the large particle contact network (approximately triangular). 
Figure \ref{config_2} 
shows a snapshot of a packing at the largest number fraction and dispersity. 
One clearly sees that the structure is a crystal with defects. 
Finally, the broad $g_{SS}(r)$ peak around $r=1$, at $f_+=0.800$, does not 
imply overlap between small particles. In fact, the $g_{SS}(r)$ value is zero 
while $r<1$. The reason for this apparent broad peak is that the bin size is 
larger for smaller average contact diameter. This quantity decreases for 
increasing number fraction and dispersity. Hence, the bin size at $f_+=0.800$ 
is significantly larger than those at lower number fractions. 

\vspace{2em}

Although not seen in the RDF plots, the area below the first peak for all 
RDFs changes as a function of the number fraction and dispersity. This might 
have implications for the mechanism 
that leads to the CPF values seen in fig. 
\ref{rpf-all}. In order to study this influence, the area below the first 
peak, in each of the RDFs, is measured. The $r$ range in which this area is 
considered is from the first peak 
position (contact diameter) up to this distance plus the dispersity, 
$\Delta R$. This is an (unnormalized) account for the average number of 
neighbors of a given type around a given particle \cite{Silbert06}. A more 
natural approach to this question would be to compare the coordination numbers 
with regard to each contact type. Any protocol for producing jammed states is 
known to produce an amount of rattlers (particles with no contacts). These 
particles also contribute to the packing relaxation. Then, a comparison of 
coordination number would exclude these particles from the analysis.
The notation corresponds to
\[
N_{SS}(\Delta R)=\int\limits_{d_{SS}}^{d_{SS}+\Delta R}g_{SS}(r)dr,
\]
\[
N_{LL}(\Delta R)=\int\limits_{d_{LL}}^{d_{LL}+\Delta R}g_{LL}(r)dr,
\]
\[
N_{SL}(\Delta R)=\int\limits_{d_{SL}}^{d_{SL}+\Delta R}g_{SS}(r)dr.
\]

This choice for the integration limits ensures that, for small-small contacts, 
only small particles are within this range. At the first few 
dispersity values, which are of the order of the $g(r)$ bin size, this 
integration leads to an overestimation of the number of neighbors. 

\vspace{2em}
\begin{figure}[h]
\rotatebox{0}{\epsfig{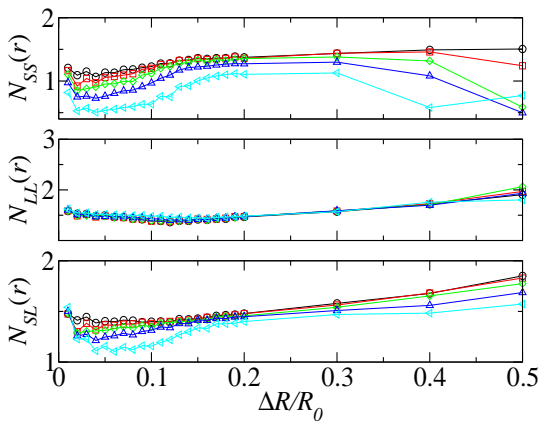}}
\caption{Area under the first peak of the pair correlations of the three 
contact types. Symbols follow fig. \ref{rpf-all}.
\label{g_area}}
\end{figure}

One can directly infer that up to intermediate dispersities, where the critical 
packing fraction reaches its lowest value, small particles increase their 
probability of being around other particles, of both types (top and bottom 
panels) with a more pronounced effect at $f_+=0.800$, while the probability for 
large-large particle contacts reaches its 
lowest value (middle panel). In addition to that, the large-large contacts 
barely change with number fraction and its minimum value occurs at a 
dispersity value close to those of the minimum jamming density, fig. 
\ref{rpf-all}.

These results, along with the RDFs in figs. \ref{rdf-1} and \ref{rdf-2}, imply 
that, less efficient packing correspond to more small particle clusters and 
less large particle ones. A possible explanation can be 
given by the fact that the initial compression is provided solely by large 
particles and the packing relaxation should take the available 
space provided by small-small neighbors, {\em i.e.}, large particles should 
push small ones in order to decrease the potential energy. This will deform the 
initial structure and small-small 
contacts will be formed in a distinct structure that the initial one. Since 
this structure is the most dense possible, the critical packing will 
occur at a lower density. Also, 
the larger the number of small particle contacts, the larger is the space 
available for structure rearrangements. Therefore, less decompression cycles 
will be needed to reach the minimum energy.


\vspace{2em}
\begin{figure}[h]
\rotatebox{0}{\epsfig{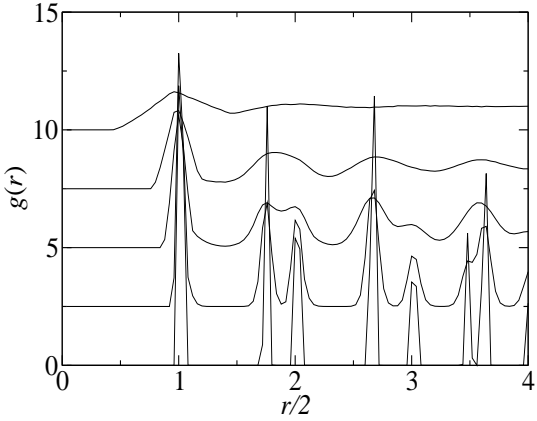}}
\caption{Pair correlation of multidisperse packings for dispersity values 
(from top to bottom): $\Delta R/R_0=0.500$, $0.200$, $0.120$, $0.050$ and 
$0.010$. Curves are shifted for clarity.
\label{rdf_multi}}
\end{figure}

For polydisperse packings, the jammed structure shows a completely distinct 
scenario. Since there are several particle sizes, the chance for the formation 
of crystalline regions during the packing relaxation is very low. Therefore, 
the packing structure should be strongly amorphous, as shown in fig. 
\ref{rdf_multi}, at large $\Delta R/R_0$. Moreover, this amorphization seems 
to be a continuous process, since the peaks at low dispersity become smoother 
for larger $\Delta R/R_0$ until they merge and, eventually, disappear.

\begin{figure}[h]
\rotatebox{0}{\epsfig{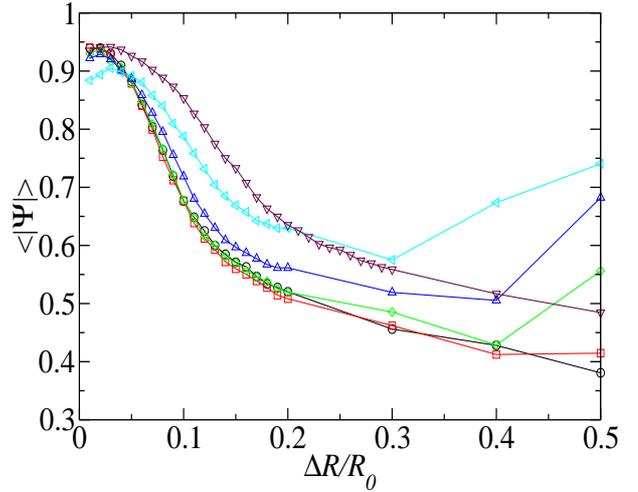}}
\caption{Orientational order parameter for all number fractions and dispersity 
values. Symbols and colors are as in fig. \ref{rpf-all}: $f_+=0.400$ 
(circles), $f_+=0.500$ (squares), $f_+=0.600$ (diamonds), $f_+=0.700$ 
(triangles), $f_+=0.800$ (left triangles), polydisperse (inverted triangles).
\label{psi_all}}
\end{figure}

Since the results for the RDFs implied that the small particle contacts 
introduced disorder, one can find a correlation between the CPF and the 
contacts orientational order. Fig. \ref{psi_all} has the average value, over 
particles and runs, measured at the end of the decompression, of the 
orientational order parameter related to the triangular lattice, eq. 
(\ref{psi}). It should be noticed that, given the definition of first neighbors 
followed here, this order parameter does not contain any contribution from 
rattlers.
\vspace{2em}

First of all, one see that, at low dispersity, this order parameter decreases 
continuously. This is consistent with the arguments given earlier, that small 
particle contacts introduce disorder in the system and it packs less 
efficiently.
Moreover, the fast increase of the critical packing fraction at large 
dispersities and number fractions can be seen to correlate with a fast increase 
in the orientational order, also consistent with the appearance of large 
crystalline regions. 
On the other hand, the smooth increase in $\left<\phi\right>$ with dispersity 
is not accompanied with a corresponding increase in the order parameter.
This implies that the inversion in the CPF curves are due to pockets of large 
particle clusters, as seen in fig. \ref{g_area} (middle panel), regardless of 
the overall decrease in order. Large particles clusters have a high value of 
$\left|\left<\Psi\right>\right|$, and, given the smooth increase in $N_{LL}(r)$ 
with dispersity, one may infer that their number is small, since the whole 
system has a low value of the order parameter. Only when large particle 
clusters are formed, the global order increases, providing a more dense packing.




In the polydisperse case, the orientational order parameter continuously 
decreases with increasing dispersity. Its value 
 is higher than for the bidisperse cases up to $\Delta R/R_0=0.200$. This is 
surprising for one would expect that more particle sizes would lead to less 
order (from eqs. (\ref{mom-uni} and (\ref{mom-uni-2})) one can see that the 
size dispersion of a polydisperse packing is proportion to $\Delta R$). Given 
that there is no monodisperse limit for this packing when 
$\Delta R/R_0\rightarrow1$, one can imagine that the order parameter for the 
$f_+=0.400$ and $0.500$ cases will eventually be higher than the polydisperse 
one at some dispersity beyond $0.500$.


\section{\label{summary}Summary and conclusions}
It was presented a numerical study on the jamming properties and structure of 
a two dimensional packing of elastic disks, for bi- and polydisperse cases. The 
attention was focused on 
the value of the maximal packing fraction for which the compression energy is 
zero. This was measured through numerical decompression experiments of a 
disordered packing, initially arranged in a crystal (triangular lattice) 
structure. The critical packing fraction (CPF) was measured at the end of the 
decompression and was shown as a function of the dispersity degree, 
$\Delta R/R_0$. Also, for bidisperse packings, the CPF was also studied as a 
function of the number fraction of large disks.

The general trend of the CPF is initially decreasing up to a minimum value, 
and then increasing with dispersity, fig. \ref{rpf-all}. The lowest CPF value 
is close to the RCP value, but obtained for a number fraction of $f_+=0.400$, 
instead of the original value obtained at $f_+=0.500$ \cite{Her02}. 
The distinct CPF behavior with $\Delta R/R_0$ observed in 
\cite{Xu09,Schreck10} is due to the fact that, in the 
present case, one starts from the most possible dense packing and, therefore, 
the introduction of disorder, through the quench and internal structure 
rearrangement, will certainly lead to a lower jamming packing fraction, since 
a more efficient packing can be achieved with a increase in local order 
\cite{Tor00}. 

At low dispersity, the system behaves approximately as the monodisperse 
crystal, as seen in the results for the average number of decompression steps, 
fig. 
\ref{avg-rad}. The departure from the monodisperse regime can be attributed to 
an increase in the number of small-small particle contacts, fig. \ref{g_area}.

The structure reveals that, for low dispersity, the decompressed packing has 
significant order, as revealed by the long range behavior of $g_{SS}(r)$, 
$g_{LL}(r)$ and $g_{SL}(r)$. 
The packing structure is mostly disordered at intermediate dispersities, and at 
the highest dispersity, large-large particle contacts bear most of the 
translational order in the system, which forms a crystal with defects, with 
the few small particles scattered between the scarce space available between 
the large particle contacts.

For polydisperse packings the long range order is completely 
absent for large dispersities, since a disk size is chosen from a uniform 
distribution of values in the range $[-\Delta R, \Delta R]$, for larger 
$\Delta R$, there will a very broad distribution of particle sizes.

The data for the local orientational order gives a similar picture. For larger 
number fractions, the packings are more ordered locally compared to lower 
number fraction cases. Also, for low dispersities, the packing is more 
ordered, as expected since it behaves like the monodisperse one. Above the 
low dispersity range, the local orientational order decreases with 
$\Delta R/R_0$, regardless of the smooth increase in the CPF. However, the 
fast increase of the CPF at large number fraction and dispersities 
is strongly correlated with a fast increase in this order parameter, implying 
that such packings are close to a crystal.

\section*{Acknowledgements} I thank C. Brito for a very welcome reading of this 
paper. This work is financially supported by CNPq and FAPESPA.

\end{document}